\documentclass[subscriptcorrection,upint,varvw,mathalfa=cal=euler,balance,hyphenate,french]{asmejour} %
\usepackage{soul}
\usepackage{siunitx}
\usepackage{mathtools, cuted}
\usepackage{flushend}
\usepackage{lipsum}
\usepackage{dblfloatfix}

\hypersetup{%
	pdfauthor={John H. Lienhard},                       		   	
	pdftitle={ASME Journal Paper LaTeX Template},                  	
	pdfkeywords={ASME journal paper, LaTeX template, BibTeX style, asmejour class},
	pdfsubject = {Describes the asmejour LaTeX template},			
	pdflicenseurl={https://ctan.org/pkg/asmejour},
}

\JourName{Mechanical Design}

\begin{document}


\SetAuthorBlock{Shammo Dutta }{
Department of Mechanical Engineering,\\
The University of Alabama,\\
Tuscaloosa, AL 35487, USA \\
email: sdutta5@crimson.ua.edu} 

\SetAuthorBlock{Girish Krishnan}{%
Industrial and Systems Engineering,\\
University of Illinois at Urbana-Champaign, \\
Urbana, Illinois 61801, USA \\
email: gkrishna@illinois.edu
}
\SetAuthorBlock{Sree Kalyan Patiballa \CorrespondingAuthor}{%
Department of Mechanical Engineering,\\
The University of Alabama, \\
Tuscaloosa, AL 35487, USA \\
email: spatiballa@ua.edu
}   
\title{DRAFT: A Design Framework for Compositional Hierarchical Mechanical Metamaterials via a Qualitative Unit-Cell Library}



\begin{abstract}
Hierarchically designed mechanical metamaterials involve nested levels of structural organization, mimicking natural structures (such as bones, wood, and bird feathers) to create advanced functional materials. Compositional hierarchy, a specific type of hierarchical strategy that involves the methodical assembly of discrete building blocks, offers unique advantages in engineering design due to its modular nature. This involves proper selection and spatial arrangements of distinct microstructures, as a result of which the desired macro-scale mechanical behavior can be achieved. Towards the design of such compositional hierarchical metamaterials, this paper presents a two-step design framework. First, material optimization of the design domain is performed using a parameterized elasticity matrix to obtain optimal conceptual designs. Second, building-block microstructure geometries are selected from a qualitative library and subjected to shape–size refinement to satisfy the desired kinematic or stiffness requirements. To construct the qualitative library, a novel parametrization scheme is initially introduced, which categorizes the planar orthotropic elasticity matrix into four distinct classes. Utilizing a kinetostatic load flow visualization technique, the candidate microstructure geometries are then populated within these four classes. The framework is validated for the design of a cantilever beam with a specified lateral stiffness requirement and the design of planar sheets that exhibit specified target deformation patterns. Thus, the present work provides a systematic and physically intuitive methodology applicable to arbitrary kinematic deformation and stiffness requirements.
\end{abstract}

\date{\today}

\maketitle 

\section{Introduction}  \label{Intro}
\noindent Metamaterials are materials that are engineered to possess extremal properties, often not found in nature. The global properties of such materials are governed by their microstructural geometry rather than their chemical composition \cite{zadpoor2016mechanical}. Mechanical metamaterials are a class of metamaterials with nonintuitive mechanical properties such as negative Poisson's ratio \cite{Bertoldi2010,lakes1987foam,patiballa2020design}, negative thermal expansion coefficient \cite{Wang2016,sigmund1997design,lakes2007cellular}, negative compressibility \cite{nicolaou2012mechanical,grima2012three,grima2008truss}, snap under tension \cite{rafsanjani2015snapping}, high modulus \cite{kadic2012practicability,huang2011topological,xia2015design}, etc. Mechanical metamaterials that undergo deformations have potential applications in soft robotics \cite{rafsanjani2019programming,mark2016auxetic}, shape morphing mechanisms \cite{shaw2019computationally,wagner2017large}, wearable devices \cite{konakovic2016beyond,papadopoulou2017auxetic,ali2014auxetic}, stretchable electronics \cite{shintake2019sensitivity,lee2019auxetic,jiang2018auxetic} and devices that adapt according to their environment \cite{mirzaali2018shape,kaur2019toward}.

Design complexity of a mechanical metamaterial can be measured using the number of independent design variables required to define a single microstructure. Including  multiple scales in the design of mechanical metamaterials, increases the geometric complexity significantly while enhancing the overall mechanical behavior. Such inclusion of multiscale design leads to hierarchical mechanical metamaterials (HMMs) that has recently become popular in the field of material and structural design \cite{Lakes1993}. Hierarchy, as such, is a universal phenomenon observed in the complex biological and natural systems. From the nanoscale collagen fibril organization in bone to the microscale platelet arrangement in nacre, biological materials achieve extraordinary macroscale mechanical properties through strategic hierarchical design across multiple length scales. This architectural approach where structure, not merely the chemical composition, dictates function — has inspired a paradigm shift in engineered material design. Hierarchy in biological and natural systems are broadly classified into four categories: porous, morphological, structural and compositional hierarchy \cite{ChenNSR}. Porous hierarchical systems comprises of multi-modal porous structures with a varying porosity across scales (e.g. trabecular bone structure \cite{Gibson_Ashby_1997}). Systems with morphological hierarchy possess a multi-level microstructures with geometric form variations (e.g. bird feathers \cite{Prum,feather}). Structural hierarchy is observed in system comprising of repetitive arrangement of structural elements (e.g. collagens in tendon \cite{Fratzl2007}). Compositional hierarchical systems possess a strategically assembled arrangement of distinct microstructures (e.g., nacre's brick-and-mortar aragonite-biopolymer architecture \cite{Wegst2014}).   

Among these hierarchical strategies, compositional hierarchy, where discrete building blocks are strategically assembled into functional systems, offers unique advantages for engineering design due to its modular nature that enables systematic exploration of the design space. This involves proper selection and spatial arrangements of distinct microstructures, as a result of which, desired macro-scale mechanical behavior can be achieved. Few simple examples of such hierarchy are observed in Rubik's cube, LEGOs, cellular materials and alloys. Realizing the potential of compositional hierarchy in engineered systems requires structured materials where: (i) discrete building blocks with distinct mechanical properties can be identified, (ii) these units can be spatially arranged in arbitrary patterns, and (iii) local unit interactions produce predictable macroscale behavior. Hierarchical mechanical metamaterials, whose global properties emerge from its microstructural geometry and their spatial arrangement, satisfy these requirements and have become a prominent platform for compositional hierarchy in mechanical design. 

\begin{figure*}[t]
\begin{center}
\includegraphics[width=1.0\linewidth]{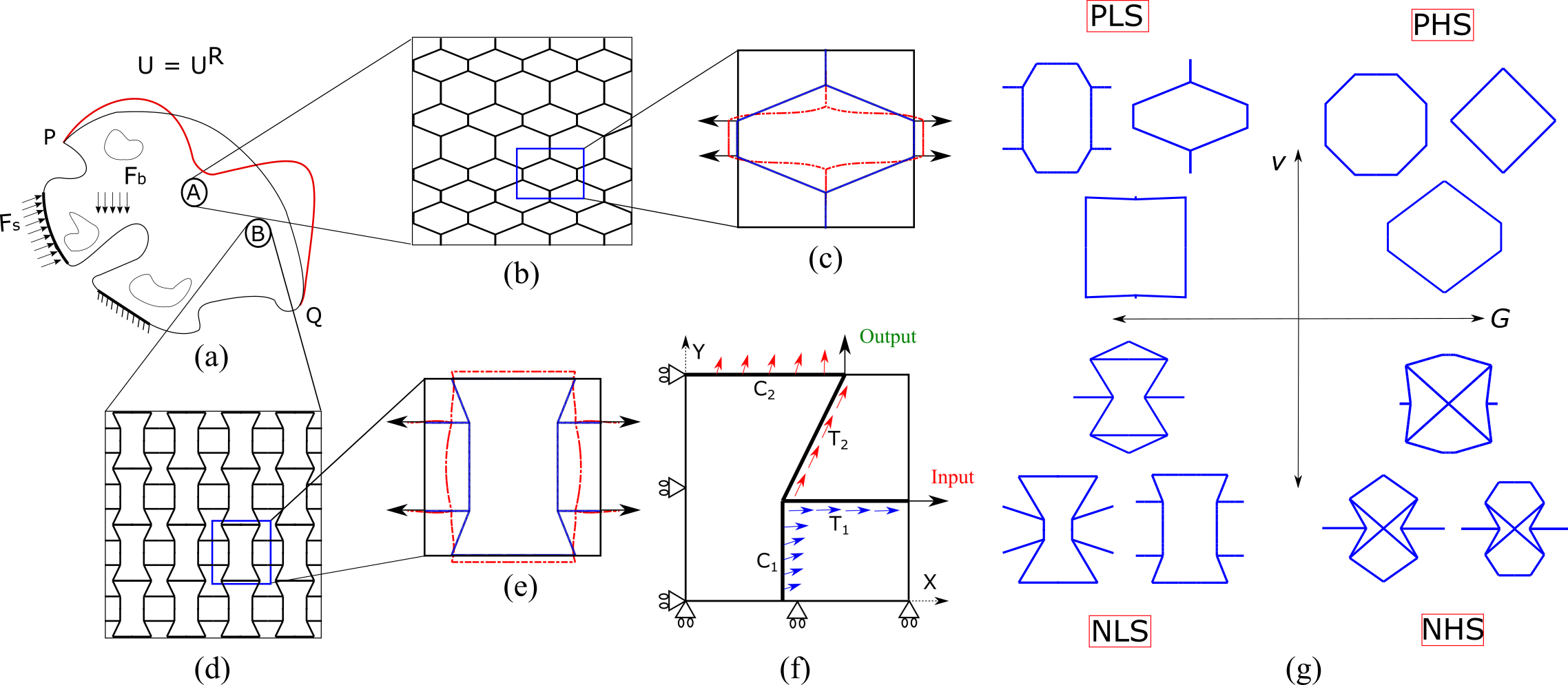}
\end{center}
\caption{Problem definition for design of compositional hierarchical mechanical metamaterials (a) design domain with given boundary, loading conditions and desired deformation (shown in red) along an edge $PQ$ (b,c)  microstructural array at location $A$ and micro-scale deformation of the corresponding single microstructure to achieve the desired macro-scale deformation (d,e) microstructural array at location $B$ and micro-scale deformation of the corresponding single microstructure to achieve the desired macro-scale deformation (f) Load flow visualization of a quarter microstructure with transmitters and constraints (g) Qualitative classification of the design space into Positive Poisson's ratio with low shear (PLS) and high shear (PHS), Negative Poisson's ratio with low shear (NLS) and high shear (NHS)}
\label{Figure-1}
\end{figure*}
Research in the field of mechanical metamaterials has mainly focused on the structural design for desired extreme mechanical properties. Along with such designs, involving hierarchy in the metamaterial geometry leads to a vast design space exploration and improved controllability of the bulk material behavior. Despite the advantages and wide potential, design of compositional HMMs is challenging as it requires considering the topology and deformations at different hierarchical levels, as shown in Fig. \ref{Figure-1}. The crux of the problem is to determine the topology of the microstructural level of the hierarchy that enforces the global geometry to deform in a predetermined fashion. This design problem can be depicted as in Fig. \ref{Figure-1}(a) with different loading conditions, boundary conditions, and desired macro-scale deformation $U^{R}$ (shown in red) along the edge $PQ$ of the domain. In such a context, different microstructures are to be designed across the domain with varying micro-scale deformations that synchronously lead to the desired macro-scale deformation. For example, a positive Poisson's ratio microstructural array (Fig. \ref{Figure-1}(b)) with corresponding micro-scale deformation as in Fig. \ref{Figure-1}(c) will achieve the required macro-scale deformation at location $A$. Similarly, a negative Poisson's ratio microstructural array at location $B$ (Fig. \ref{Figure-1}(d)) with corresponding micro-scale deformation as in Fig. \ref{Figure-1}(e) will lead to the desired macro-scale deformation at that location. Furthermore, the design is challenging as there are no existing design methodologies that systematically provide insights into the design solutions while delegating the tedious task of optimizations to a computer.

Due to their nonintuitive behaviors and inherent scale complexities, which lead to functionally enhanced mechanical properties, the design of compositional HMMs has been an active field of research in materials design. For instance, introducing a hierarchical truss-based design in a rotating auxetic metamaterials resulted in reduced stiffness due to a drastic change in the deformation mechanism (rotation to concurrent rotation and flexure), which in turn resulted in a less localized stress concentration \cite{Mizzi2020}. Similarly, just using hierarchical multi-level rotating squares, auxetic behavior and porosity can be controlled to a larger extent \cite{Gatt2015}. Combining the concept of rotating unit auxetic systems and cubic crystal arrangements resulted in hierarchical rotating mechanical metamaterials exhibiting high stiffness/density ratio and giant negative Poisson's ratio \cite{Mizzi2025}. Coupling vertex-based structural hierarchy with disordered microstructure resulted in superior energy absorption under predefined stress constraints \cite{Zhu2025}. Such rational design methods, though intuitive, are difficult to generalize to any design problem. Furthermore, additional dimensions, due to hierarchy, results in a large amount of design variables across multiple scales. 

To tackle the complexities involved in design of HMMs, robust computational methods such as topology optimization (TO) and data-driven design are often adopted. TO-based design metamaterials using inverse homogenization traces back to the seminal work done by Sigmund \cite{sigmund1994materials,sigmund1995tailoring}. Classical TO-based design methods involves computational schemes such as the solid isotropic material with penalization (SIMP), evolutionary structural optimization (ESO), and level-set method (LSM). Based on these schemes, recently various methods were proposed to design HMMs. Gao et al. \cite{Gao2019} designed a compositional cellular HMM using a free-material distribution optimization method to get the initial optimized density distribution, then a concurrent TO scheme using parametric level set method (PLSM) is used to obtain the optimized macrostructure and material microstructures. Ai et al \cite{Ai2019} adopted a similar PLSM-based TO method in combination with strain energy-based homogenization and meshfree algorithm to design 2-D compositional cellular HMMs. Huo et al. \cite{Huo2025} formulated a unique approach to design a three scale functionally graded hierarchical metamaterial utilizing a bi-directional homogenization scheme, where a forward scheme from micro to mesoscale and an inverse homogenization scheme from macro to mesoscale is adopted. Along with these typical computational design methods, various data-driven design methods in-combination with TO schemes have also garnered significant interest \cite{Zheng2021,Wang2021,Wang2020}. While the aforementioned computational methods are mathematically robust, they are time-consuming, sensitive to the algorithm used, and the obtained solutions may require extensive postprocessing to be practical. In some cases, the process does not guarantee manufacturable design solutions and decouples user from the design process and may yield limited user insight. 
\begin{figure*}[t]
\begin{center}
\includegraphics[width=1.0\linewidth]{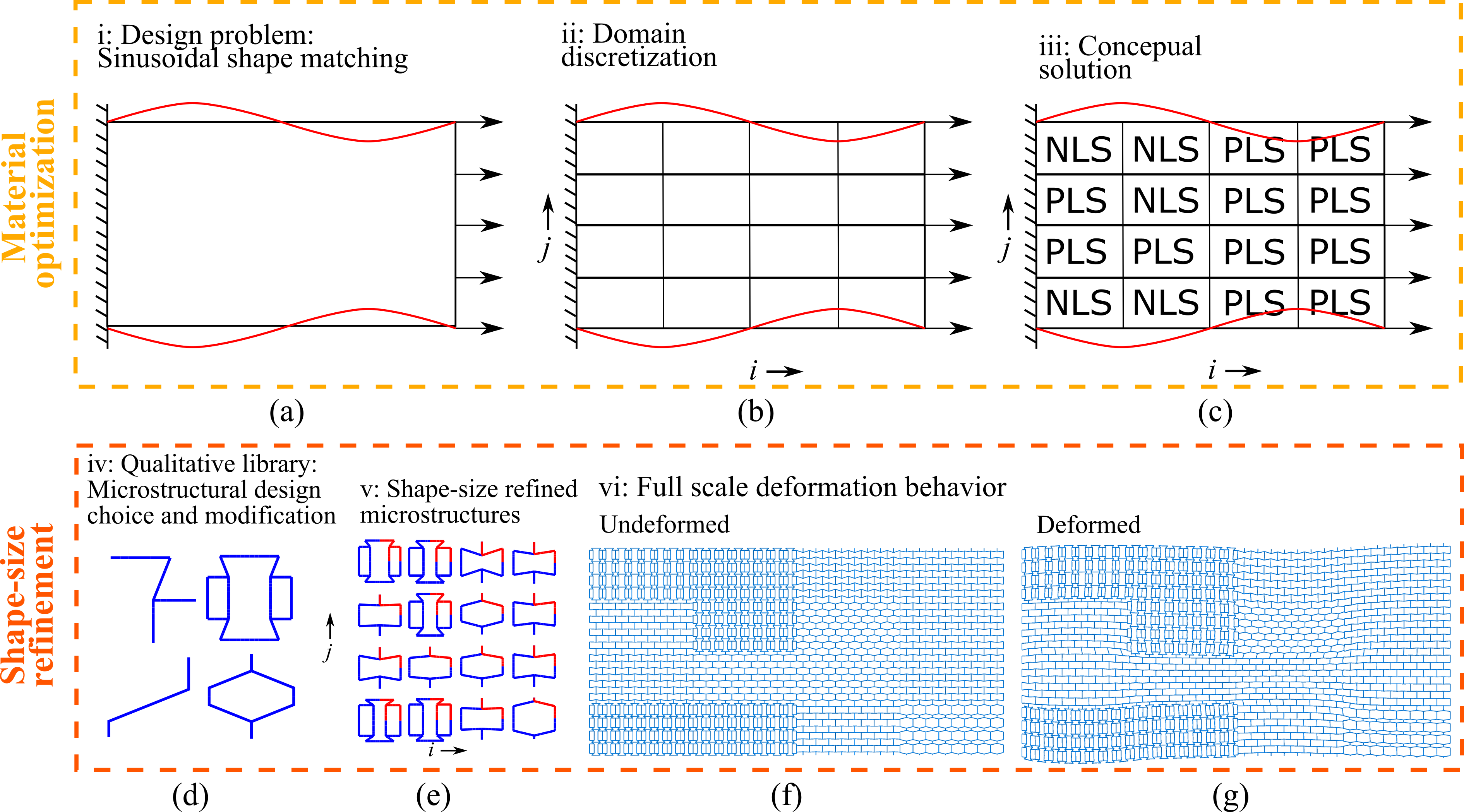}
\end{center}
\caption{Complete two-step framework for the design of compositional hierarchical mechanical metamaterials : (a) Start with a design problem, perform material optimization: (b) domain discretization and (c) conceptual design solution, (d) choose conceptual solution from the qualitative library, (e) perform shape-size refinement to obtain (f)-(g) the final solution.}
\label{Figure-2}
\end{figure*}

Focusing on the aforementioned limitations and prioritizing designer insight during design process, we propose an alternative two-step design framework for compositional hierarchical mechanical metamaterials (HMMs) driven by a qualitative library of metamaterial microstructures. First phase of this work includes the formation of a qualitative library of microstructures based on their mechanics. Followed by this, we introduce the first step, where we obtain optimal conceptual designs by performing material optimization on the design domain with a parameterized elasticity matrix. The obtained conceptual solutions are chosen from the developed qualitative library of feasible microstructures. In the second step, a shape-size refinement where the preselected microstructure geometry is fine-tuned to exactly meet the desired kinematic or stiffness requirements. This process allows the designer to control the creative part of the design process, while the more tedious shape-size refinement is performed in the second step. The present work thus provide a systematic and physically intuitive methodology applicable to arbitrary deformation and stiffness requirements. A preliminary version of this work was presented at the ASME IMECE 2020 conference \cite{PatiballaIMECE}. The current article significantly expands on our conference work by incorporating detailed guidelines for the two-step design of hierarchical mechanical metamaterials, analysis of mesh discretizations and periodic unit cell array sizes on the final design solution (Figs. 5 and 6), and additional experimental results (Figs. 4 and 10). \\ 
\indent The paper is organized as follows. Section 2 presents the qualitative library of the planar microstructure geometries built using load flow visualization and parametrization of the elasticity matrix. Section 3 lays out the guidelines for the systematic two-step design process for compositional HMMs. Section 4 presents various shape-matching examples and an application of compositional HMMs to design a flexible door-latch mechanism. Certain examples are showcased to validate the proposed design framework both numerically and experimentally. Finally, we summarize the contributions, conclusions, and future work in Section 5. 

\section{Qualitative library of planar mechanical metamaterials} \label{SEC:Ch6_Sec1}
\noindent Planar microstructures are the building blocks, which, when repeated as in Fig. \ref{Figure-1} leads to a functional metamaterial. While there could be infinitely many such microstructures, we show in this section that they can be qualitatively classified into a finite set, which greatly simplifies the design space exploration. This results in a qualitative library of planar mechanical metamaterials, which is formed on the basis of load flow visualization technique.

\subsection{Load flow visualization} \label{Subsec : LFV}
\noindent Load flow is a vector field of fictitious forces that flow through the mechanism geometry from input to output. These fictitious forces called ``transferred loads" can be quantified using the compliance matrices that relate forces and displacement between any two points in the topology, as detailed in \cite{krishnan2013kinetostatic,Patiballa2018}. In a compliant mechanism, load flow enables functional characterization of the mechanism geometry into transmitters and constraints. A member with a predominantly axial load flow is called a transmitter, while a member with transverse load flow and moment components can be called a constraint. This framework can be used to analyze a mechanism by decomposing it into building blocks of transmitter-constraint sets \cite{krishnan2013kinetostatic}. In Fig. \ref{Figure-1}(f), a negative Poisson's ratio microstructure along with the load flow visualization from the input point to the output, is presented. As the microstructure has orthotropic symmetry, we consider only a quarter of the microstructure. The quarter microstructure has two transmitters $T_{1}$, $T_{2}$ with predominantly axial load flow, and two constraints $C_{1}$,$C_{2}$ with transverse load flow. The decomposition of a mechanism topology based on load flow into transmitters and constraints provides a qualitative understanding of the the mechanism's functionality. Hence, load flow visualization enables qualitative analysis of mechanism topology. This understanding can be inverted for conceptual design synthesis. This was demonstrated for the design of planar metamaterials in \cite{Patiballa2018}, design of compliant mechanism with embedded actuators in \cite{patiballa2019load}, design of spatial - compliant and shape morphing mechanisms in \cite{patiballa2018conceptual,krishnan2020conceptual}, and design of three-dimensional mechanical metamaterials in \cite{patiballa2020design}. 

\begin{figure*}[t]
\begin{center}
\includegraphics[width=1.0\linewidth]{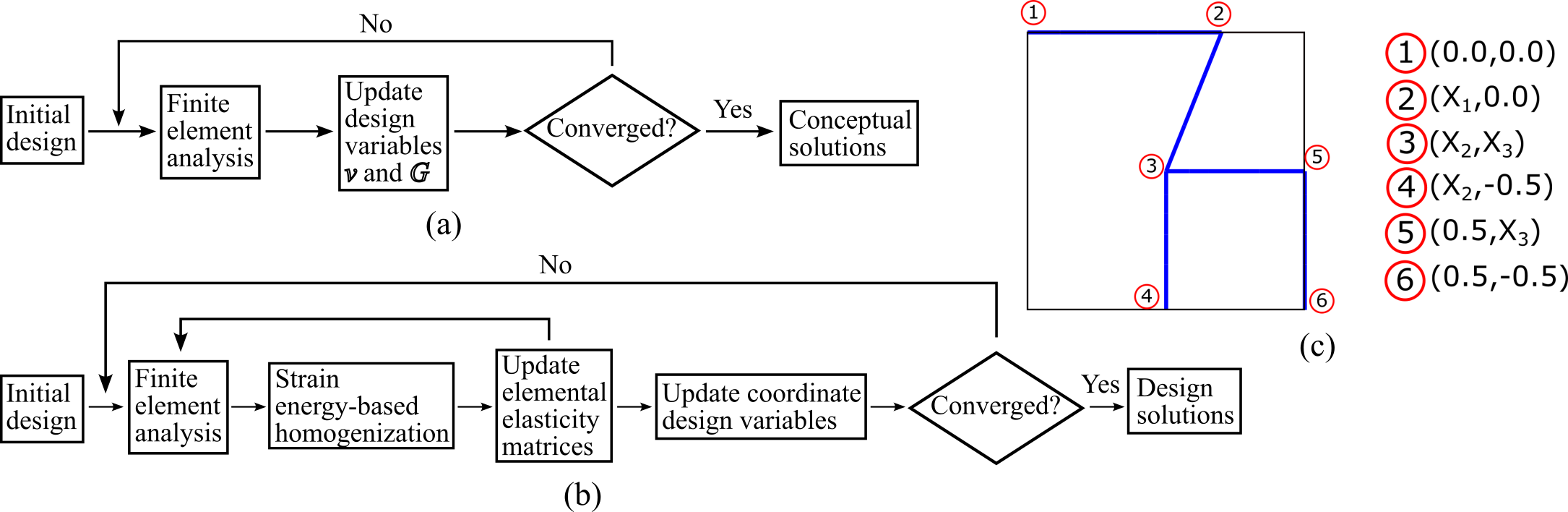}
\end{center}
\caption{Process flow for (a) material optimization (b) shape-size refinement and (c) coordinate design variables used in shape-size refinement of the conceptual geometry}
\label{Figure-3}
\end{figure*}
\subsection{Qualitative classification of planar metamaterials using Load flow visualization (LFV)} \label{Subsec : Qual-Lib}

\noindent Based on the nature of load flow in the microstructure, a qualitative analysis and a classification scheme for orthotropic auxetic microstructures presented in reference \cite{Patiballa2018}. These insights are generalized for both positive and negative Poisson's ratio microstructures below

\begin{enumerate}[label=(\roman*)]
    \item A material with negative Poisson's ratio demonstrates reversal in load flow direction such that any applied strain in positive $X$ (or $Y$) direction is transferred as a corresponding force in the positive $Y$ (or $X$) direction.
    \item A material with positive Poisson's ratio does not posses such a reversal in load flow behavior. 
    \item A material with high shear modulus demonstrates conflict of load flow orientation in their transmitters due to input strain in different faces.
    \item A material with low shear modulus does not posses significantly conflicting load flow orientation in their transmitters. 
\end{enumerate}

These guidelines qualitatively classify the entire orthotropic planar material space into four classes : 

\begin{enumerate}
    \item Positive Poisson's ratio with low shear (PLS)
    \item Positive Poisson's ratio with high shear (PHS)
    \item Negative Poisson's ratio with low shear (NLS)
    \item Negative Poisson's ratio with high shear (NHS)
\end{enumerate}

This classification facilitates a qualitative library of all the possible microstructures in the planar case, which can be seen in Fig. \ref{Figure-1} (g). Figure \ref{Figure-1} (g) depicts only a few examples for each of the classes. More designs can be obtained by following the load flow-based design methodology presented in \cite{Patiballa2018}. This qualitative library offers several alternative solutions to the microstructure selection problem. It is important to note here that the qualitative library is not just limited to the microstructures presented in Fig. \ref{Figure-1} (g). One can apply the aforementioned design insights to obtain more microstructures, which will further expand the qualitative library. 

\subsection{Parameterizing the elasticity matrix}
\noindent The design of compositional hierarchical mechanical metamaterials entails determining material microstructures that make up the global topology to meet a given kinematic or stiffness requirement. In this paper, we investigate a sequential approach where we first determine the microstructure's qualitative class and then refine a selected solution from the database to match the exact problem specifications. 

We have seen that the two important factors for classification in planar mechanical metamaterials are Poisson's ratio ($\nu$) and shear stiffness ($G$). A general orthotropic elasticity matrix in planar metamaterials is shown in Eq. \ref{Ch6_Eq1}. There are four independent components in the elasticity matrix i.e, $C_{11}$, $C_{12}$, $C_{22}$ and $C_{33}$. Each component represents an elastic stiffness in different loading conditions. $C_{11}$ is an elastic stiffness in $X$-direction, $C_{22}$ is an elastic stiffness in $Y$-direction, $C_{12}$ is a bi-axial stiffness, and $C_{33}$ is shear stiffness. Poisson's ratio is the ratio between $C_{12}$ and $C_{11}$. By dividing the elasticity matrix by $C_{11}$, we obtain Eq. \ref{Ch6_Eq2}. Here, $C_{22}$/$C_{11}$ is the ratio of elastic stiffness in $Y$ direction to elastic stiffness in $X$ direction. {Since, this is not a parameter that can be classified qualitatively, we assume that the elastic stiffness in $X$ and $Y$ direction is equal.} But, $C_{22}$/$C_{11}$ is taken as a value slightly different from 1.00 for numerical stability and to preserve the positive definiteness of the elasticity matrix. Now, the elasticity matrix can be qualitatively parameterized as in Eq. \ref{Ch6_Eq3}. 

\begin{equation}
      C=\begin{bmatrix}
   {{C}_{11}} & {{C}_{12}} & 0  \\
   {{C}_{12}} & {{C}_{22}} & 0  \\
   0 & 0 & {{C}_{33}}  \\
\end{bmatrix} \label{Ch6_Eq1}\\ 
\end{equation}
\begin{gather}
    C = \begin{bmatrix}
   {{C}_{11}}/{{C}_{11}} & {{C}_{12}}/{{C}_{11}} & 0  \\
   {{C}_{12}}/{{C}_{11}} & {{C}_{22}}/{{C}_{11}} & 0  \\
   0 & 0 & {{C}_{33}}/{{C}_{11}}  \\
\end{bmatrix} 
= \begin{bmatrix}
   1.00 & \nu & 0  \\
   \nu & {{C}_{22}}/{{C}_{11}} & 0  \\
   0 & 0 & G^*  \\
\end{bmatrix}
\label{Ch6_Eq2}
\end{gather}

This parameterized elasticity matrix can lead to our four qualitative classes according to the values of Poisson's ratio and shear modulus. Classification with respect to Poisson's ratio is straightforward as the value can either be greater or lower than zero. In comparison, the classification of shear modulus is challenging because it can either be high or low. We consider normalized shear stiffness values that are more than 0.5 as high and less than 0.5 as low. To push the values close to the extremes, we penalize the shear modulus similar to solid isotropic material with penalization (SIMP) \cite{bendsoe1999material} with penalization factor $p = 3$. 

\begin{equation}
     C=\left[ \begin{matrix}
   1.00 & \nu  & 0  \\
   \nu  & 1.01 & 0  \\
   0 & 0 & G^{p}  \\
\end{matrix} \right] 
\label{Ch6_Eq3} 
\end{equation}

This parametrized elasticity matrix is used in each of the discretized elements in the design domain. This parametrization allows for classification of the each element into one of four qualitative classes from Fig. \ref{Figure-1} (g). For example, if $\nu > 0$ and $G ~ 1$, then this element is a positive Poisson's ratio with high shear. 

\section{Systematic two-step design framework} \label{SEC: Ch6_design}
\noindent In this section, we present a systematic two-step framework for the design of compositional hierarchical mechanical metamaterials, as depicted in Fig. \ref{Figure-2}. The first step is a material optimization on the design domain with a parameterized elasticity matrix to determine which qualitative class each microstructure belongs. In the second step, a microstructure is chosen from the qualitative library, and a shape-size refinement is performed on the preselected microstructure geometry to meet desired kinematic or stiffness requirements. 
 
\subsection{Material optimization step}
\noindent Here, we present the design guidelines with the aid of an example shown in Fig. \ref{Figure-2}(a). The design problem is to deform a planar sheet so that it can expand into a sinusoidal shape along its boundaries perpendicular to the loading direction. The desired deformation is shown in red. This is an example of a kinematic requirement, where the planar sheet has to meet the desired shape. 

\begin{itemize}
\item \textit{Step I}: Setup a design domain and discretize the domain into elements. 
\end{itemize}

\noindent In the first step, setup a design domain with applied loads, desired deformations, and boundary conditions. Then, discretize the design domain into various quadrilateral plane stress elements, as in Fig. \ref{Figure-2}(b). Here, each element has a different parameterized elasticity matrix. The discretization in this example is arbitrary, but including more number of elements can give more accurate deformations at the expense of increasing computational cost. For this example, we have chosen a discretization of 4 $\times$ 4 elements in $X$ and $Y$ directions. A detailed discussion on the choice of discretization is given in Section \ref{params-choice}.

\begin{itemize}
\item \textit{Step II}: Perform a material optimization to find the qualitative classes of each of the elements.
\end{itemize}

\begin{align}
  & \underset{{{X}_{i}}}{\mathop{\min }}\,:{{\sum_j^n \left  ( U_j -{U_j}^{*} \right)}^{2}} \label{Ch6_Eq4} \\ 
 & s.t: Lb_{i} \leq X_{i} \leq Ub_{i} \label{Ch6_Eq5} 
\end{align}

\noindent Now, material optimization can be performed to obtain the conceptual, qualitative classes of each of the design elements. The objective of the optimization in this example is to match the desired displacements. The objective function is posed as a least-squares minimization, as depicted in Eq. \ref{Ch6_Eq4}. Here, $U_j$ is the displacement of the $j^{th}$ node in the design domain, and $U_j^{*}$ is the desired displacement at that point. The design variables are $X_{i}$'s, which are the Poisson's ratio and shear modulus of each of the elements.

The process flow for material optimization is shown in Fig. \ref{Figure-3}(a). We initialize the design with elemental elasticity matrices with Poisson's ratio of $0.5$ and shear modulus of $0.5$. Then, we iteratively run the finite element analysis to update the Poisson's ratio and shear modulus values of each of the elements until convergence. Once the convergence is achieved, we classify the elemental elasticity matrices into one of the four qualitative classes presented in Section \ref{SEC:Ch6_Sec1}. For the classification of each elemental elasticity matrix, we use a threshold of 0 for Poisson's ratio and 0.5 for shear modulus. So, if $\nu > 0$ corresponds to positive Poisson's ratio and $\nu <0$ corresponds to negative Poisson's ratio. In the case of shear modulus, if $G > 0.5$, we consider the elasticity matrix to be high shear, and $G < 0.5$ corresponds to low shear. It is important to note that the solution obtained here is conceptual in nature and not unique. The optimization is performed in MATLAB using the ``fminunc" function. For this problem, the conceptual solution is shown in Fig. \ref{Figure-2}(c). Here, the Poisson's ratio varies from negative to positive depending upon the peaks of the target sinusoidal shape. 

\subsection{Shape-size refinement step}
\noindent Once the conceptual solutions are obtained, we perform a shape-size refinement of the chosen microstructure from the qualitative library to meet the desired kinematic or stiffness requirements. 

\begin{itemize}
\item \textit{Step III}: Choose the microstructural designs from the qualitative library.  
\end{itemize}

\noindent In this step, we chose the microstructural designs pertaining to each of the elements' conceptual, qualitative solutions. The conceptual solution gives a combination of positive and negative Poisson's ratio with low shear (PLS and NLS). Now, we chose these classes from the qualitative library shown in Fig. \ref{Figure-2}(d). {The size of the chosen microstructural unit cell is 1$mm$ $\times$ 1$mm$}. And shape-size refinement is performed only on a quarter of the unit cell, as shown in Fig. \ref{Figure-2}(d).

\begin{itemize}
\item \textit{Step IV}: Modify or re-design the material microstructure to ensure connectivity between the elements.  
\end{itemize}

\noindent One of the challenges in choosing microstructures from the library and periodically filling the elements is the connectivity between the different elements. To ensure the connectivity between the elements, we can either modify the existing design or re-design the microstructure using load flow-based design, as detailed in \cite{Patiballa2018,Patiballa2017}. Here, we modify the existing PLS and NLS microstructure design, as shown in Fig. \ref{Figure-2}(d). We modify the design by adding an extra beam to ensure the connectivity between the elements.

\begin{itemize}
\item \textit{Step V}: Perform a shape-size refinement of the chosen microstructural designs to meet the desired deformations.  
\end{itemize}

\noindent Next, we perform a shape-size refinement of the chosen microstructure geometry to meet the desired deformations. The objective of the geometrical optimization is similar to the material optimization of Eq. \ref{Ch6_Eq4}. Here, $U_j$ is the displacement of the $j^{th}$ node in the design domain, and $U_j^{*}$ is the desired displacement at that point. But, the design variables $X_{i}$'s are the coordinates of the points, as shown in Fig. \ref{Figure-3}(c). The thickness of the beams is unchanged in this formulation. 

\begin{figure}[h]
\begin{center}
\includegraphics[width=1.0\linewidth]{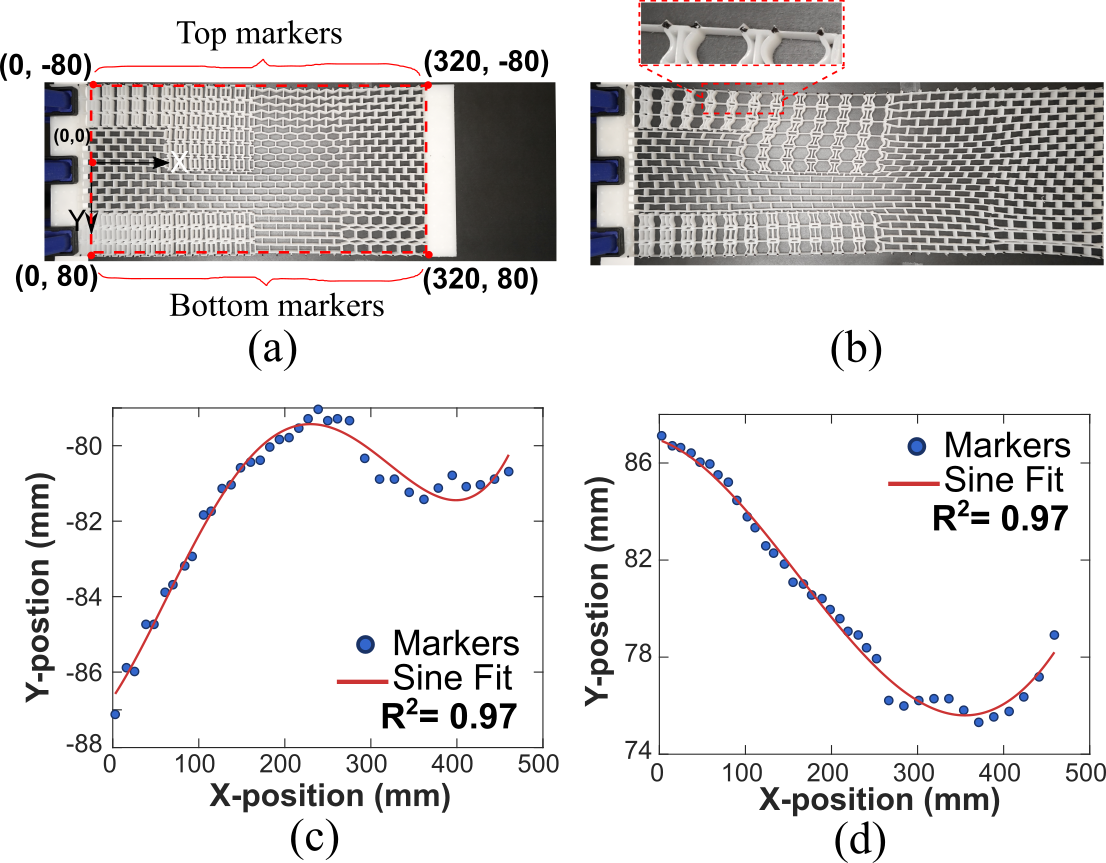}
\end{center}
\caption{Qualitative validation of the design framework for the sinusoidal shape matching}
\label{Figure-4}
\end{figure}

\begin{figure*}[t]
\begin{center}
\includegraphics[width=1\linewidth]{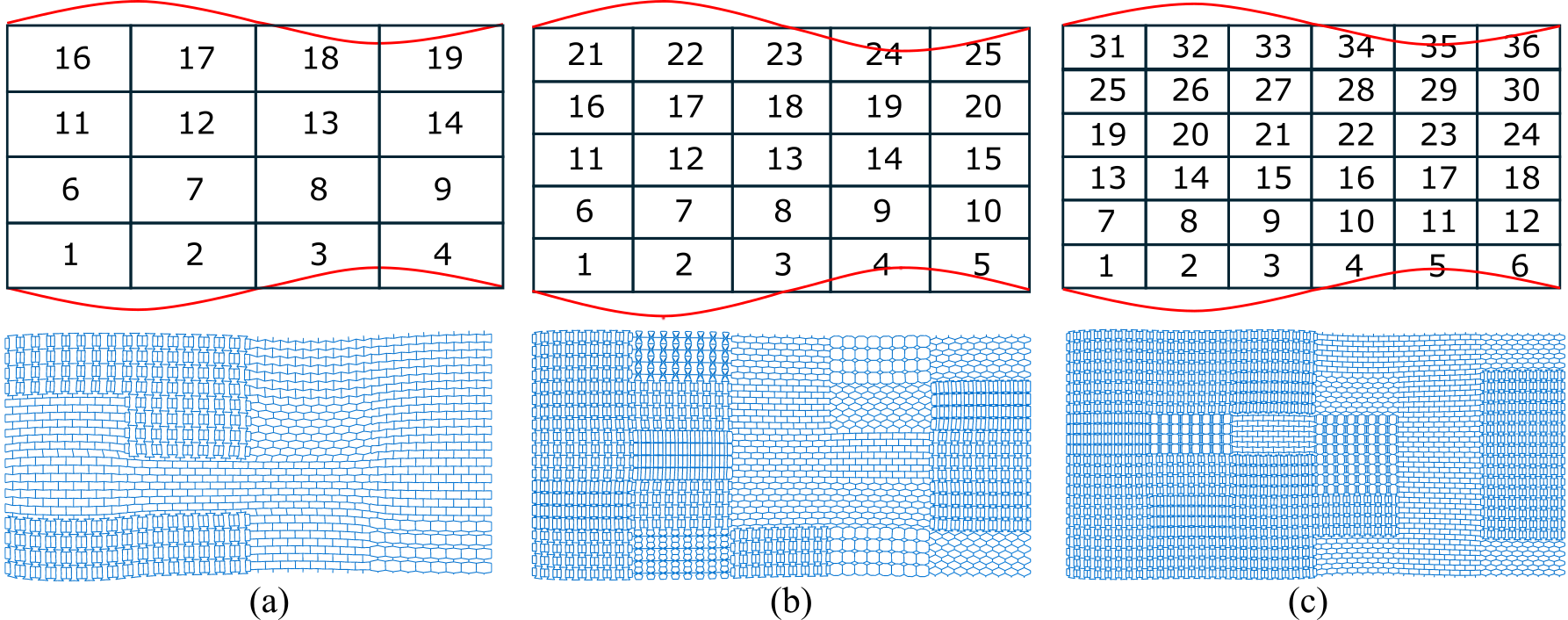}
\end{center}
\caption{Effect of design domain discretization on the sinusoidal shape-matching problem: (a) 4 $\times$ 4 (b) 5 $\times$ 5 (b) 6 $\times$ 6 }
\label{var-disc}
\end{figure*}

\begin{figure*}[b]
\begin{center}
\includegraphics[width=1\linewidth]{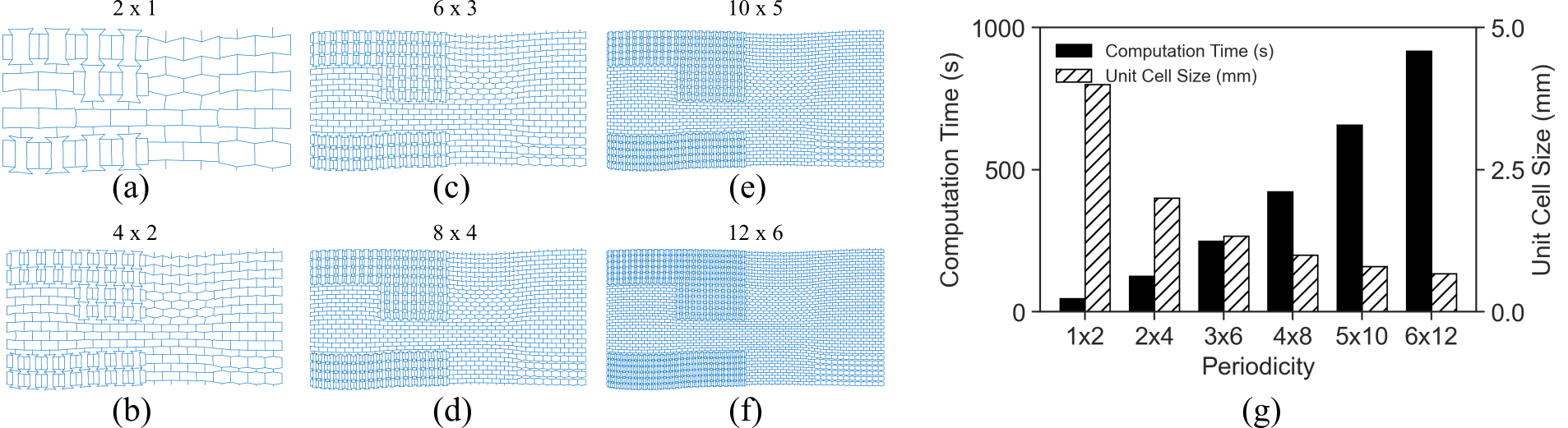}
\end{center}
\caption{Effect of periodic unit cell array size for a 4 $\times$ 4 discretized and optimized sinusoidal shape-matching metamaterial: (a-f) 2 $\times$ 1, 4 $\times$ 2, 6 $\times$ 3, 8 $\times$ 4, 10 $\times$ 5, 12 $\times$ 6, and (g) Computation time (solid bars) and unit-cell size (hatched bars) for varying periodicity levels }
\label{var-period}
\end{figure*}
The overall process flow of the shape-size is shown in Fig. \ref{Figure-3}(b). We initialize the design with coordinates of the microstructure. Then, we compute the homogenized elasticity matrix in each of the elements using strain energy-based homogenization (refer to \cite{zhang2007using,Patiballa2018} for details). The homogenized elasticity matrices in each of the elements are used to compute the global stiffness matrix in the finite element analysis. We, iteratively, run this process by updating the design variables until convergence. This procedure runs a macro optimization to meet the desired deformations while running a shape-size optimization of the microstructures. The final geometry of the microstructures obtained for each of the sixteen elements can be seen in Fig. \ref{Figure-2}(e).  

\begin{itemize}
\item \textit{Step VI}: Perform a finite element analysis to validate the solution. 
\end{itemize}

\noindent In this final step, we perform a nonlinear finite element analysis on the full scale design obtained after shape-size optimization to validate the design framework. Here, we use a commercial software package ABAQUS to run the nonlinear finite element analysis using 2-node linear beam elements with hybrid formulation (B21H). A hyperelastic material is used for the analysis with the Neo-Hookean model. The constants $C_{10}$ and $D_{1}$ are \SI{0.106}{\mega\pascal} and \SI{0.03}{\mega\pascal^{-1}} respectively for all the examples in this paper \cite{mirzaali2018shape}. The initial configuration of the design domain is shown in Fig. \ref{Figure-2}(f) and the deformed configuration is shown in Fig. \ref{Figure-2}(g).  

Following the entire design process, we now experimentally observe the shape matching behavior of the compositional HMM optimized for a sinusoidal deformation. Using the optimized unit cell coordinates obtained from the sinusoidal shape-matching problem, a solid CAD model was constructed in SolidWorks. The geometry was scaled to an overall size of $320\text{mm} \times 160\text{mm}$, with each strut modeled at $1\text{mm}$ width and $10\text{mm}$ out-of-plane thickness. A negative mold was 3-D printed using PLA Basic (Bambu Lab P1S FDM printer) and Dragon Skin 10 A (Slow) silicone was used for casting the complete design, as shown in Fig.\ref{Figure-4}(a). For deformation tracking, black edge markers were applied using a custom adhesive made by mixing Smooth-On black pigment with Sil-Poxy, as shown in the inset of \ref{Figure-4}(b)-inset. The fabricated design was tested on a flat base with a black background and reference lines spaced $170\text{mm}$ apart. One end of the structure was clamped, while the other was manually pulled. Images of the deformation were captured using a top-mounted camera under controlled lighting. Marker positions were extracted using ImageJ (Fiji), and a generalized sine curve was fitted to the edge marker trajectory. Figure \ref{Figure-4}(b) shows the deformed configuration, while Fig.\ref{Figure-4}(c,d) shows the fitted curve. The results qualitatively confirm that the boundary nodes achieves a sinusoidal shape, validating the ability of the sequential design framework to produce programmed deformation.

\subsection{Choice of parameters for the design framework}\label{params-choice}
\noindent In the design process illustrated, we notice that there are various parameters that a designer has the freedom to choose. Here, we will discuss these parameters. First, the parameters that affect the material optimization are:
\begin{figure*}[t]
\begin{center}
\includegraphics[width=0.85\linewidth]{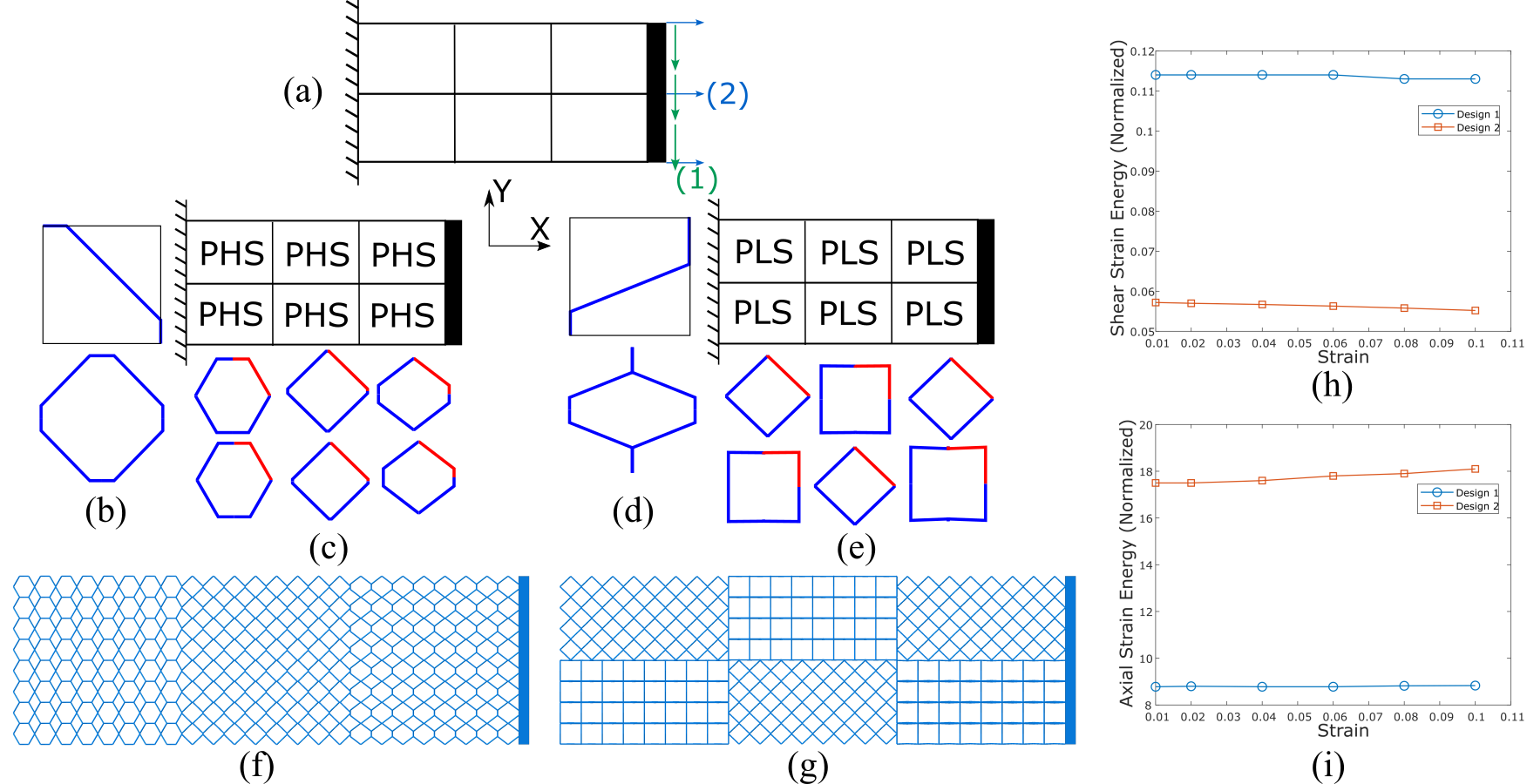}
\end{center}
\caption{Relative stiffness scenarios of the cantilever beam. (a) discretized design domain with axial and shear loading, design solution from the qualitative library, conceptual solutions from the material optimization and  final designs of each of the cells obtained from shape-size optimization (b-c) for the design problem of having high stiffness in shear loading relative to axial loading, (d-e) for the design problem of having high stiffness in axial loading relative to shear loading. Full scale designs for beam with (f) high shear stiffness and (g) high axial stiffness. Stiffness analyses of the cantilever beams showing variation of normalized strain energy (h) in shear loading with applied strain and (i) in axial loading with applied strain.}
\label{Figure-5}
\end{figure*}
\begin{enumerate}[label=(\roman*)]
    \item Mesh discretization of the design domain. 
    \item Initial guess of the Poisson's ratios and shear modulus.
\end{enumerate}

In the above example, we chose a mesh discretization of 4 $\times$ 4 elements in the $X$ and $Y$ directions. However, the designer has the freedom to choose the number of elements in the $X$ and $Y$ directions. This might affect the conceptual solution depending upon the design domain, boundary, and loading conditions. As the number of elements increases, the computational time for the material optimization increases but might provide a more accurate fit to the desired deformations (see Fig. \ref{var-disc}). Therefore, one needs to trade off between accuracy, computational cost, complex shape-matching, and having more control points. In most cases, the initial guess of the Poisson's ratio and shear modulus does not affect the conceptual solution.

The parameters that affect the shape-size refinement are:

\begin{enumerate}[label=(\roman*)]
    \item Conceptual solution chosen from the qualitative library.
    \item Periodic unit cell array size in each of the elements.
\end{enumerate}

In the example presented, we have chosen negative and positive Poisson's ratio low shear (PLS and NLS) microstructures. However, if we had chosen any other microstructure from the qualitative library, the final design solution would be different. Finally, the periodic unit cell array size in each element also affects the accuracy of the shape matching (see Fig. \ref{var-period}). Theoretically, homogenization assumes that there is an infinite periodic array of unit cells to achieve the homogenized properties. However, in reality, we choose a finite array size, and the size of the array increases the computational time of the finite element analysis in ABAQUS. For instance, increasing the periodic unit cell array size from 4 $\times$ 2 to 12 $\times$ 6 increases the computation time from nearly 120 seconds to more than 900 seconds (shown in Fig. \ref{var-period} (g)). Additionally, while constraining to a constant design domain, increasing the periodic unit cell array size in each element will significantly reduce the dimensions of the unit cell, making them impractical to fabricate using standard 3D printing methods. For example, the unit cell size reduces from 2 mm to 0.67 mm when increasing the periodicity from 4 $\times$ 2 to 12 $\times$ 6 (shown in Fig. \ref{var-period} (g)). Keeping these limitations in mind, we have chosen the periodic unit cell array size to be 8 $\times$ 4 in the $X$ and $Y$ directions, respectively.

\section{Design examples of compositional hierarchical mechanical metamaterials}
\noindent In this section, we present three design examples with different requirements to illustrate and validate the design framework of compositional hierarchical mechanical metamaterials (HMMs) presented in Section \ref{SEC: Ch6_design}. The first example is of a planar cantilever beam with desired stiffness requirements alone. The second design problem is to obtain a planar sheet which can shape change into a regular sinusoidal along one boundary edge. The third design problem is of a planar sheet with desired kinematic requirement of lateral expansion alone and, the fourth example is of a planar sheet with both kinematic and stiffness requirements. 

\begin{figure}[h!]
\begin{center}
\includegraphics[width=1.0\linewidth]{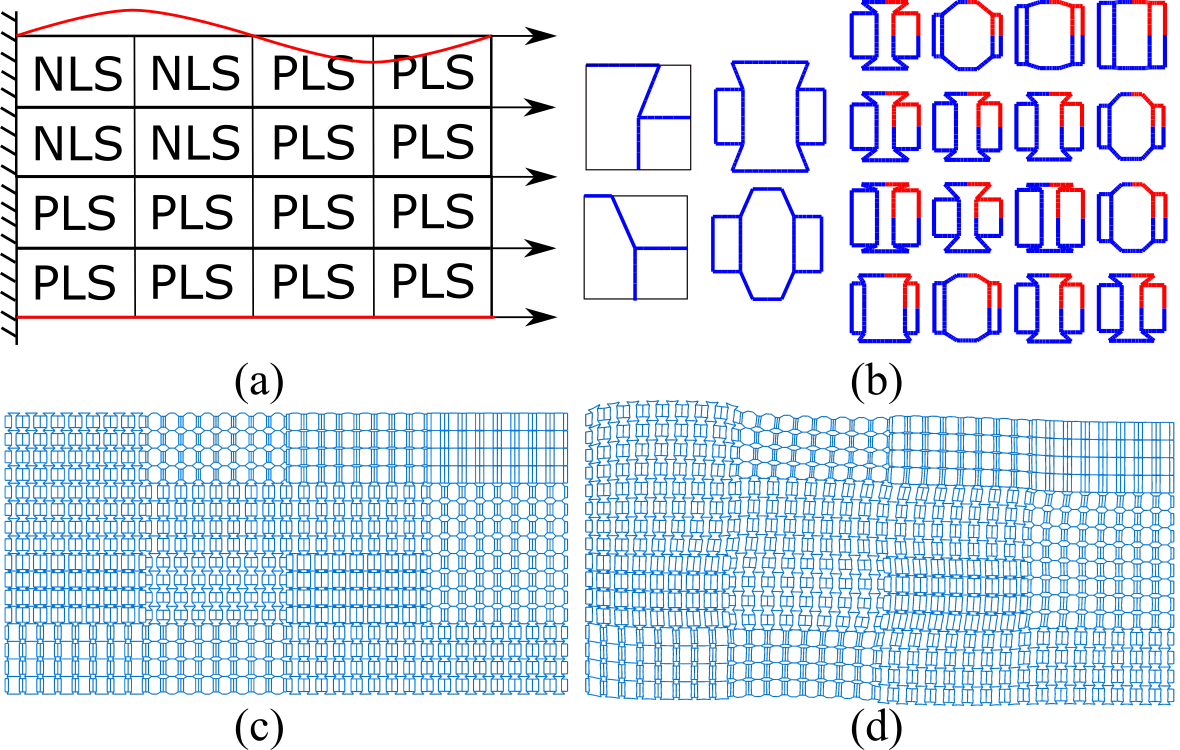}
\end{center}
\caption{Compositional hierarchical mechanical metamaterials that can shape change into a regular sinusoidal on one-edge. (a) discretized design domain with desired shape change (shown in red) and conceptual solutions from the material optimization, (b) design solution from the qualitative library and final designs of each of the cells obtained from shape-size optimization, (c,d) initial and deformed  configuration of the design in ABAQUS}
\label{Figure-6}
\end{figure}

\subsection{Cantilever beam}
\begin{figure*}[t!]
\begin{center}
\includegraphics[width=\linewidth]{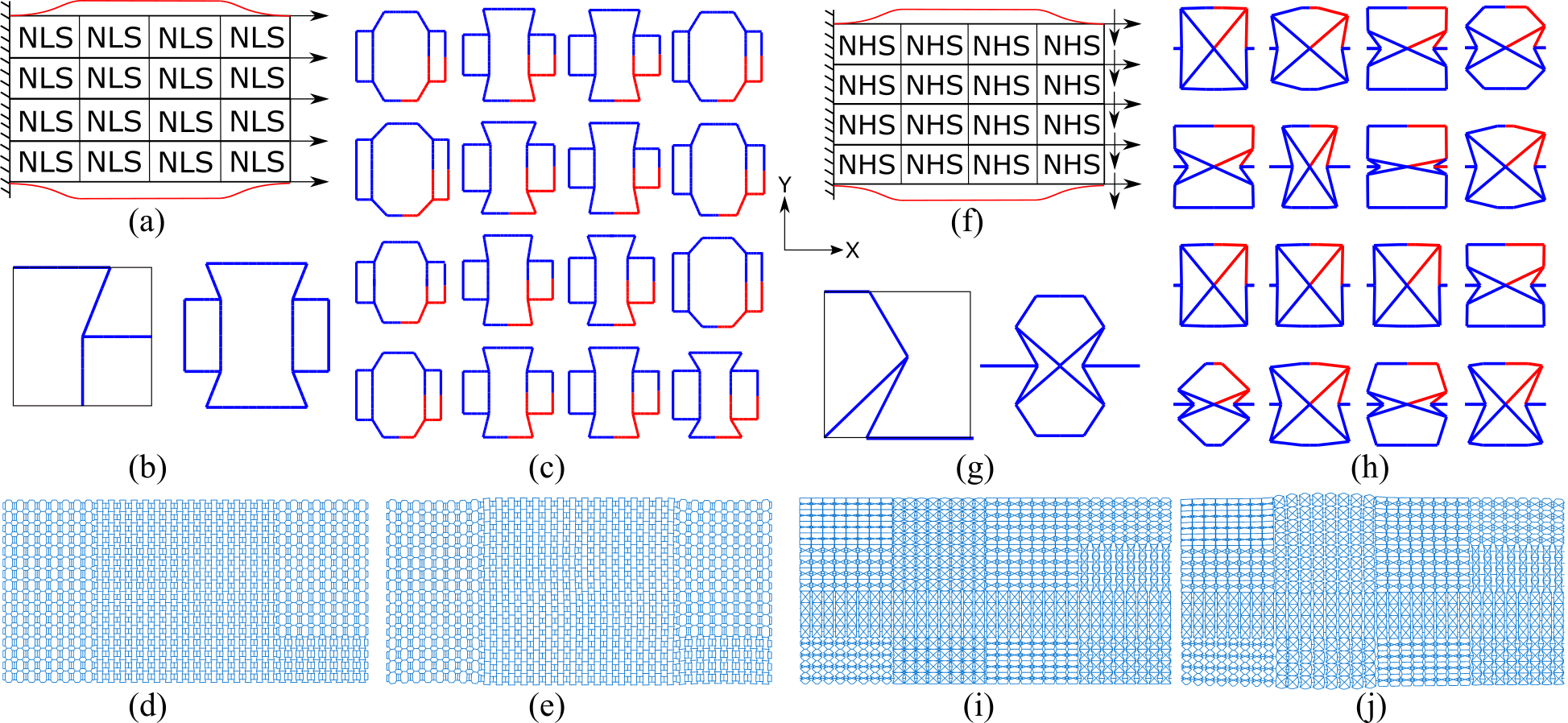}
\end{center}
\caption{Compositional hierarchical mechanical metamaterials that expands laterally (a-e) and sheet that can expands laterally and have high stiffness in $Y$-direction. (f-j). (a,f) discretized design domain with desired shape change (shown in red) and conceptual solutions from the material optimization, (b,g) design solution from the qualitative library, (c,h) final designs of each of the cells obtained from shape-size optimization, (d,i) initial configuration of the design domain in ABAQUS, (e,i) deformed configuration in ABAQUS}
\label{Figure-7}
\end{figure*}
\noindent The first example illustrated here is of a cantilever beam, where the design problem is to obtain optimal material microstructures that can maximize the stiffness of the beam in either shear loading or axial loading conditions. The design scenario is shown in Fig. \ref{Figure-5}(a), where there is axial loading in $X$-direction and shear loading in $Y$-direction. In this scenario, one design problem is finding the material microstructures that can maximize the stiffness of the cantilever beam in $Y$-direction relative to the stiffness in $X$-direction, i.e., have high stiffness in shear loading relative to axial loading. The objective function can be formulated as a ratio of the strain energies in the corresponding directions, as in Eq. \ref{Ch6_Eq6}, where (1) corresponds to shear loading, and (2) corresponds to axial loading. Following the proposed design steps in the previous section, we obtain a conceptual solution after the material optimization shown in Fig. \ref{Figure-5}(c) (top). The microstructural geometry for positive Poisson's ratio high shear (PHS) chosen from the qualitative library shown in Fig. \ref{Figure-5}(b). Next, we run a shape or geometry optimization. The final design geometries of each of the elements are shown in Fig. \ref{Figure-5}(c) (bottom). The full scale design for the same is shown in Fig. \ref{Figure-5}(f).

The second design problem in this scenario is finding the material microstructures that can maximize the stiffness of the cantilever beam in $X$-direction relative to the stiffness in $Y$-direction, i.e., have high stiffness in axial loading relative to shear loading. The objective function for this problem is shown in Eq. \ref{Ch6_Eq7}. We follow a similar procedure as in the first design problem. The conceptual solution and the chosen microstructure from the qualitative library are shown in Fig. \ref{Figure-5}(e) (top) and Fig. \ref{Figure-5}(d) respectively. The final design geometries of each of the elements after shape optimization are shown in Fig. \ref{Figure-5}(e) (bottom). The full scale design for the same is shown in Fig. \ref{Figure-5}(g).

\begin{align}
  & \underset{{{X}_{i}}}{\mathop{\min }}\,:{\frac{SE_{1}}{SE_{2}}}, 
 & s.t: Lb_{i} \leq X_{i} \leq Ub_{i} \label{Ch6_Eq6} \\
   & \underset{{{X}_{i}}}{\mathop{\min }}\,:{\frac{SE_{2}}{SE_{1}}},
 & s.t: Lb_{i} \leq X_{i} \leq Ub_{i} \label{Ch6_Eq7} 
\end{align}

Finally, to validate the relative stiffness solutions of a cantilever beam in Fig. \ref{Figure-5}(f,g), we plot the normalized strain energy variation with strain as in Figs. \ref{Figure-5}(h,i) respectively. Design 1 is for achieving a metamaterial with high stiffness in shear loading relative to axial loading, while design 2 is for achieving a metamaterial with high stiffness in axial loading relative to shear loading. From the shear strain energy plot of Fig. \ref{Figure-5}(h), we can notice that design 1 has high strain energy validating the high stiffness in shear loading. Similarly, from the axial strain energy plot of Fig. \ref{Figure-5}(i), we can notice that design 2 has high strain energy validating the high stiffness in axial loading.   

\subsection{One-sided sinusoidal shape-matching}
\noindent The second design problem illustrated here is a one-sided sinusoidal shape matching metamaterial. Here, one side of the design domain should deform as a sinusoid, while the other edge has no deformation. The design domain and conceptual solution are shown in Fig. \ref{Figure-6}(a). The initial geometries chosen from the qualitative library are shown in Fig. \ref{Figure-6}(b). The final design geometries after the shape optimization are depicted in Fig. \ref{Figure-6}(c). The finite element analysis validation of the solution is shown in Fig. \ref{Figure-6}(d). For the design examples of sinusoidal and one-sided sinusoidal shape matching, the Poisson’s ratios across a design oscillate between positive and negative values depending upon the target shapes. 

\subsection{Lateral expansion} \label{allExpanse}
\noindent The third design problem illustrated here is a shape matching compositional hierarchical mechanical metamaterials (HMMs) that can expand in a direction perpendicular to the loading direction. Following the similar design steps presented in Section \ref{SEC: Ch6_design}, we obtain the conceptual solution and the final design geometries after shape optimization. All the elements for this design problem are negative Poisson's ratio with low shear (NLS). This conceptual solution provides necessary insights for the designer. The design domain, conceptual solution, and the initial geometry from the qualitative library are shown in Figs. \ref{Figure-7}(a,b). The final design geometries are shown in Fig. \ref{Figure-7}(c). The finite element analysis validation is performed in ABAQUS. The initial and deformed configurations are shown in Figs. \ref{Figure-7}(d,e) respectively.

\subsection{Shape-matching with stiffness requirement} \label{allExpanse_stiff}
\begin{figure*}[b]
\begin{center}
\includegraphics[width=0.85\linewidth]{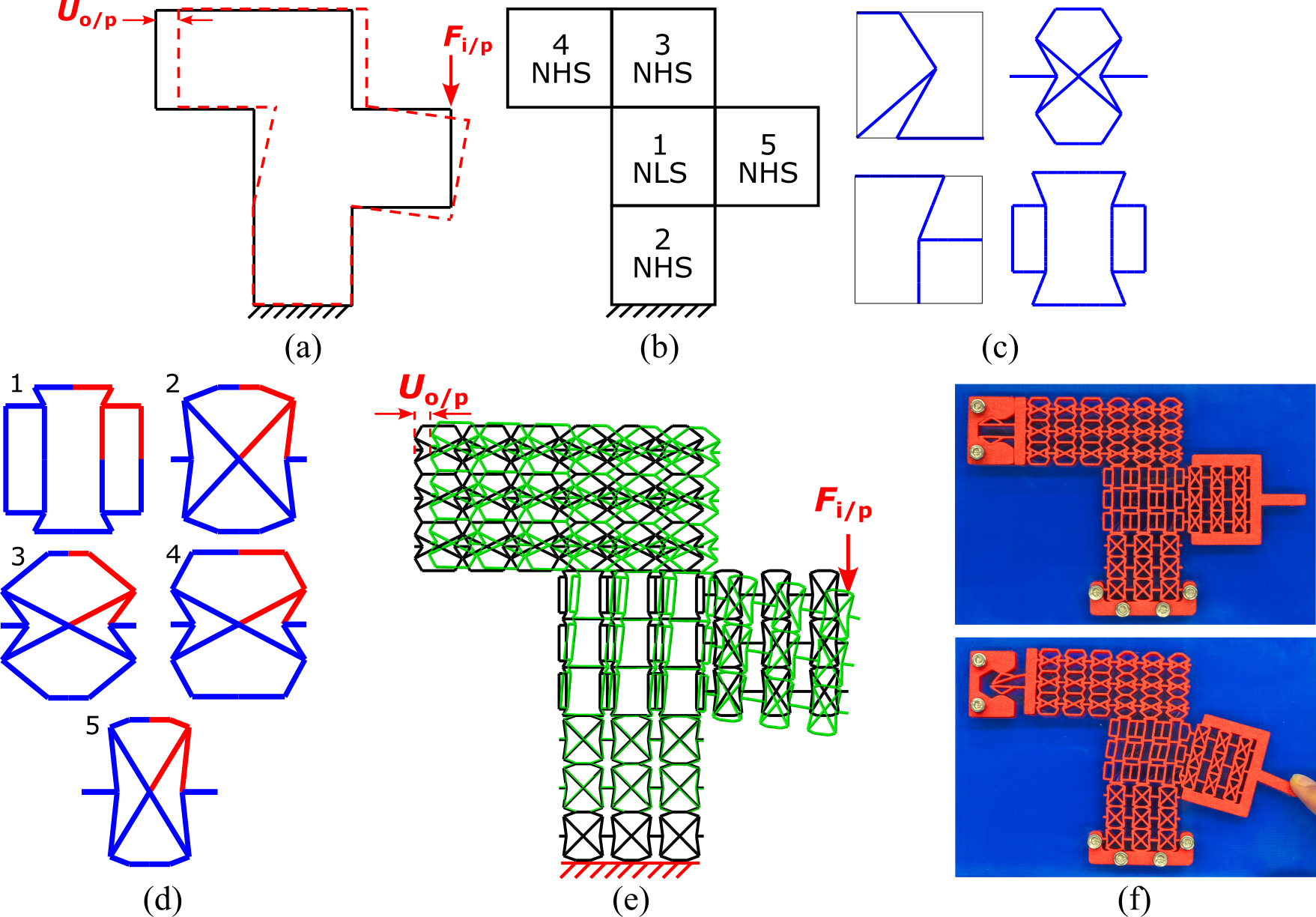}
\end{center}
\caption{Design of door latch mechanism using compositional hierarchical mechanical metamaterials (HMMs): (a) Design domain with boundary conditions and target deformation (b) domain discretization and conceptual solution obtained from material optimization step (c) choice of microstructures from the qualitative library (d) final optimized microstructures obtained from the shape-size refinement step (e) full scale finite element analysis (f) fabricated mechanism}
\label{Figure-8}
\end{figure*}
\noindent In this design example, we want to obtain compositional HMMs with both kinematic and stiffness requirements. The design problem is similar to Fig. \ref{Figure-7} (a). Here, in addition to the kinematic requirement of expansion in a direction perpendicular to the loading direction, we also need the planar sheet to be stiff in $Y$-direction. The conceptual solution is shown in Fig. \ref{Figure-7}(f). As expected, the conceptual solution gives microstructures with high shear modulus. We chose the initial geometry of Fig. \ref{Figure-7}(g) from the qualitative library. The objective function is to match the desired displacements and have high stiffness in $Y$-direction. So, the objective function is posed as a sum of least square minimization and strain energy minimization in $Y$-direction. The final design geometries after the optimization are depicted in Fig. \ref{Figure-7}(h). The solution is validated in finite element analysis, as shown in Figs. \ref{Figure-7}(i,j). From the aforementioned design examples, we infer that the proposed design optimization framework for compositional HMMs is applicable to a wide range of design problems involving both kinematic and stiffness matching properties. In the next section, we now apply this established framework to showcase the design of a door latch mechanism composed of compositional HMMs.

\section{Application of compositional hierarchical mechanical metamaterials (HMMs): Simplified door latch mechanism design}
\noindent Design framework devised for compositional HMMs in this work enables a building block-based design approach utilizing the proposed qualitative library of planar mechanical metamaterials. Using this building block design method, we showed how we can design planar sheets to meet both kinematic and stiffness requirements. In this section, we further utilize the proposed method to design a simplified door latch mechanism. The objective function for this problem is posed as a least-squares minimization, as depicted in Eq \ref{eq-8}. To ensure feasible fabrication, we considered additional distance and angular constraints between $i^{th}$ and $i+1^{th}$ points' coordinates of individual microstructures, as shown in Eq. \ref{eq-9}. The values of $l_{min}$, $\theta_{min}$, and $\theta_{max}$ were chosen based on the available 3D printer's minimum manufacturable dimensions.Figure \ref{Figure-8} (a) shows the initial design domain (solid black straight lines) with the boundary conditions and the target deformation (dashed red straight lines). 
\begin{align}
        & \min_{X_i}:\sum_{j}^{n }( U_j -{U_j})^2 \label{eq-8}\\ 
        & \text{s.t: } Lb_{i} \leq X_{i} \leq Ub_{i} & l_{i,i+1} \geq l_{min} \text{, } \theta_{min} \leq \theta_{i,i+1} \leq \theta_{max}
        \label{eq-9} 
\end{align}

The initial design domain is discretized into five elements (Fig. \ref{Figure-8}(b)) and first subjected to the material optimization step that resulted in the conceptual design solution shown in Fig. \ref{Figure-8}(b). Before even conducting the material optimization, from Fig. \ref{Figure-8}(a) we can intuitively presume that element '1' needs to shear the most while the other elements needs to shear the least. The results obtained from the material optimization, as shown in Fig. \ref{Figure-8}(b), confirms this observed presumption, further suggesting the strong intuitive nature of the proposed design framework. Followed by the material optimization, we chose the initial microstructures (Fig. \ref{Figure-8}(c)) from the qualitative library developed in Sec. \ref{Subsec : Qual-Lib} and conducted shape-size refinement step to obtain the final optimized microstructures, as shown in Fig. \ref{Figure-8}(d). The full scale finite element analysis (conducted in ABAQUS) reveals the latch-unlocking motion at the output region (see Fig. \ref{Figure-8}(e)). To further corroborate the obtained results from the design framework, we fabricated and tested the optimized full scale design. We 3D printed the latch mechanism using TPU material (68D shore hardness) using the Bambulab P1S FDM printer. The fabricated mechanism is shown in Fig. \ref{Figure-8} (f) (top) where the output region is modified to include a snap-fit member to capture the locking-unlocking motion and the input region is modified to include a lever. In the deformed state, followed by the input load at the lever, the microstructures in element block '1' undergoes extreme shearing while the others resists shearing deformation, resulting in unlocking of the snap-fit member, as shown in Fig. \ref{Figure-8} (f) (bottom). This highlights the versatile potential of the proposed design and optimization framework for compositional HMMs in areas spanning from structures to soft robotics.


\section{Conclusion and Future Work}

\noindent Natural and biological systems provide several strategies, such as porous, morphological, structural, and compositional, that lead to material hierarchies. Among these strategies, compositional hierarchy, where discrete building blocks are tessellated strategically to attain a global behavior, offers an excellent bio-inspiration for engineering design due to its modular nature. Building on this premise, we introduce a two-step sequential design framework to synthesize compositional hierarchical mechanical metamaterials (HMMs) with either desired kinematic-matching and/or stiffness-matching properties, based on a qualitative library of planar mechanical metamaterials. The insights from load flow visualization (LFV) and strain energy-based homogenization (SEBH) are used to construct the design framework. In-depth analysis of negative and positive Poisson's ratio microstructures using LFV provides crucial design insights that correlate the nature of load flow in a microstructure to its Poisson's ratio and shear modulus. For instance, negative Poisson ratio microstructures showed a reversal in load flow direction such that any applied strain in the positive X (or Y) direction results in a transferred force in the positive Y (or X) direction. Likewise, high shear microstructures depict a conflicting load flow orientation in their transmitter struts. Based on such guidelines and design correlations, we develop a qualitative library of planar orthotropic microstructures, which possess a combination of positive/negative Poisson's ratios and low/high shear moduli. Material optimization on the parametrized elasticity matrix leads to conceptual solutions. The conceptual solutions are taken from the qualitative library, and shape-size refinement is performed to meet the desired stiffness and/or kinematic requirements. Detailed guidelines for the two-step design framework for stiffness-matching and shape-matching compositional HMMs have been presented. Various examples are presented to demonstrate and validate the proposed design framework. It is important to note that the design framework leads to multiple solutions based on the choice of parameters such as mesh discretization, preselected microstructure geometry from the qualitative library, and periodic unit cell array size.

In future work, we will investigate an extension of the design framework to 3D compositional HMMs. Given the spatially varying nature of material properties across an optimized design domain, a potential avenue for future work could be to optimize the stress distribution using established design optimization methods \cite{DuttaJMDStress, DuttaIMECE}. Furthermore, the current design framework is sequential in nature, but to attain a more optimal framework, we will extend the framework to concurrent designs. The current framework is based on beam-based metamaterials, and this could be further embedded in a graph-based neural network model to speed up the computational cost.

\section*{Acknowledgment} 
\noindent The authors acknowledge the financial support from the National Science Foundation through grants NSF-2344385 and NSF-2418355.











\appendix   
\selectlanguage{english} 



\bibliographystyle{asmejour}   




\end{document}